\documentclass[12pt]{article}
\usepackage[dvips]{graphicx}
\usepackage{amsfonts,amsthm,amstext,amscd}
\usepackage{amssymb}
\usepackage{yfonts}
\newcommand{\beqa}{\begin{eqnarray}}
\newcommand{\eeqa}{\end{eqnarray}}

\def\d{{\rm d}}
\def\tr{\,{\rm Tr}}

\newcommand{\be}{\begin{equation}}
\newcommand{\ee}{\end{equation}}
\newcommand{\beq}{\begin{equation}}
\newcommand{\eeq}{\end{equation}}
\newcommand{\bea}{\begin{eqnarray}}
\newcommand{\eea}{\end{eqnarray}}

\newcommand{\rc}{\nonumber\\}
\newcommand{\bear}{\begin{eqnarray}}
\newcommand{\eear}{\end{eqnarray}}
\begin{document}
\baselineskip=15.5pt
\pagestyle{plain}
\setcounter{page}{1}

\def\r{\rho}   
\def\CC{{\mathchoice
{\rm C\mkern-8mu\vrule height1.45ex depth-.05ex
width.05em\mkern9mu\kern-.05em}
{\rm C\mkern-8mu\vrule height1.45ex depth-.05ex
width.05em\mkern9mu\kern-.05em}
{\rm C\mkern-8mu\vrule height1ex depth-.07ex
width.035em\mkern9mu\kern-.035em}
{\rm C\mkern-8mu\vrule height.65ex depth-.1ex
width.025em\mkern8mu\kern-.025em}}}

\def\cb{{\cal B}}
\def\cc{{\cal C}}
\def\cph{{\bold \varphi}}
\def\ch{{\cal H}}
\def\cp{{\cal P}}
\def\cst{c_T}
\def\csx{c_X}
\def\csr{c_R}
\def\css{c_S}
\def\csf{c_F}
\def\csu{c_U}
\def\cz{{\cal Z}}
\def\pperp{{\perp\perp}}
\def\to{\rightarrow}
\newcommand{\qn}{\textswab{q}}
\newcommand{\wn}{\textswab{w}}

\def\ht{{\hat{t}}}
\def\hxm{{\hat{x}^\mu}}
\def\hxn{{\hat{x}^\nu}}
\def\hr{{\hat{r}}}
\def\ha{{\hat{a}}}
\def\hb{{\hat{b}}}
\def\hc{{\hat{c}}}
\def\hd{{\hat{d}}}
\def\htau{{\hat{\tau}}}
\def\hi{{\hat{x}^i}}
\def\hj{{\hat{j}}}
\def\hK{{\hat{K}}}
\def\hL{{\hat{L}}}
\def\hM{{\hat{M}}}
\def\hN{{\hat{N}}}
\def\d{\partial}
\def\med{\frac{1}{2}}

\newfont{\namefont}{cmr10}
\newfont{\addfont}{cmti7 scaled 1440}
\newfont{\boldmathfont}{cmbx10}
\newfont{\headfontb}{cmbx10 scaled 1728}
\renewcommand{\theequation}{{\rm\thesection.\arabic{equation}}}
\font\cmss=cmss10 \font\cmsss=cmss10 at 7pt
\par\hfill KUL-TF-09/27

\begin{center}
{\LARGE{\bf Hydrodynamics of fundamental matter}}
\end{center}
\vskip 10pt
\begin{center}
{\large 
Francesco Bigazzi $^{a}$, Aldo L. Cotrone $^{b}$, Javier Tarr\'\i o  $^{c}$.}
\end{center}
\vskip 10pt
\begin{center}
\textit{$^a$ Physique Th\'eorique et Math\'ematique and International Solvay
Institutes, Universit\'e Libre de Bruxelles; CP 231, B-1050
Bruxelles, Belgium.}\\
\textit{$^b$  Institute for theoretical physics, K.U. Leuven;
Celestijnenlaan 200D, B-3001 Leuven,
Belgium.}\\
\textit{$^c$ Departamento de  F\'\i sica de Part\'\i culas, Universidade de Santiago de Compostela and Instituto Galego de 
F\'\i sica de Altas Enerx\'\i as (IGFAE); E-15782, Santiago de Compostela, Spain.}\\

{\small fbigazzi@ulb.ac.be, cotrone@itf.fys.kuleuven.be, tarrio@fpaxp1.usc.es}
\end{center}

\vspace{15pt}

\begin{center}
\textbf{Abstract}
\end{center}

\vspace{4pt}{\small \noindent 
First and second order transport coefficients are calculated for the strongly coupled ${\cal N}=4$ SYM plasma coupled to massless fundamental matter in the Veneziano limit.
The results, including among others the value of the bulk viscosity and some relaxation times, are presented at next-to-leading order in the flavor contribution.
The bulk viscosity is found to saturate Buchel's bound.
This result is also captured by an effective single-scalar five-dimensional holographic dual in the Chamblin-Reall class and it is suggested to hold, in the limit of small deformations, for generic plasmas with gravity duals, whenever the leading conformality breaking effects are driven by marginally (ir)relevant operators. This proposal is then extended to other relations for hydrodynamic coefficients, which are conjectured to be universal for every non-conformal plasma with a dual Chamblin-Reall-like description. Our analysis extends to any strongly coupled gauge theory describing the low energy dynamics of $N_c\gg1$ D3-branes at the tip of a generic Calabi-Yau cone.
The fundamental fields are added by means of $1\ll N_f\ll N_c$ homogeneously smeared D7-branes. 
}
\vfill

\newpage

\section{Introduction}
\setcounter{equation}{0}
Hydrodynamic models provide a fairly accurate description of the large-time evolution after thermalization of the QCD plasma produced at RHIC \cite{arsene}.
In this respect, a first principle computation of the transport coefficients is both relevant and challenging, due to the strongly coupled nature of the system.
Surprisingly, some features of the known transport coefficients appear to be common to many plasmas of strongly coupled theories, including QCD.
The prototype example is the shear viscosity, whose ratio with the entropy density has, in theories with two-derivative gravity duals, a universal value which is quite close to the QCD one \cite{Policastro:2001yc}.
This observation justifies a careful analysis of first and second order transport coefficients in the class of four dimensional quantum field theories admitting a gravity dual.

Hydrodynamic studies by means of the gauge/gravity correspondence have been focusing to-date on theories without fundamental flavors (apart from what concerns the universal shear viscosity \cite{kss,Mateos:2006yd,Casero:2006pt,Bertoldi:2007sf,d3d7plasma}). 
Fundamental flavors are clearly relevant in the RHIC plasma \cite{arsene,Muller:2006ee}.
In this paper, the results for the flavor contribution to the bulk viscosity and to some relevant second order transport coefficients (relaxation times, $\kappa$, $\kappa^*$) of the strongly coupled ${\cal N}=4$ SYM plasma are presented. 

These results are easily extended to ${\cal N}=1$ plasmas describing the thermal low energy dynamics of $N_c\gg1$ D3-branes at generic Calabi-Yau cones over five-dimensional compact Sasaki-Einstein manifolds $X_5$ (the ${\cal N}=4$ SYM case corresponds to $X_5=S^5$). These are conformal plasmas without fundamental fields. Flavors are added by means of $N_f\gg1$ homogeneously smeared D7-branes extended along the radial direction and wrapping compact three-manifolds $X_3$ in the space transverse to the D3-branes \cite{Bigazzi:2005md,Casero:2006pt,benini,d3d7plasma}. We will just focus on the case in which all the flavors are massless; the related non-Abelian flavor symmetry group is explicitly broken into a product of Abelian factors, due to the smearing. The addition of massless flavors induces a breaking of conformal invariance at the quantum level. 

The reason to analyze these theories stems from the fact that, despite being plagued by a Landau pole in the UV, they are the simplest examples where thermal flavor effects at strong coupling can be reliably studied, providing the first manageable toy models for the QCD plasma in its near conformal regime.

Our results are obtained at next-to-leading order in a perturbative expansion in $\epsilon_h\sim\lambda_h N_f/N_c$ where $\lambda_h$ is the 't Hooft coupling at the energy scale fixed by the plasma temperature. The precise coefficients defining $\epsilon_h$ depend on the model and hence on the volumes of the $X_5$ and $X_3$ spaces. For the ${\cal N}=4$ case, for example
\begin{equation}
\epsilon_h=\frac{1}{8\pi^2} \lambda_h \frac{N_f}{N_c}\,.
\end{equation}
Within this perturbative approach, the flavors can be considered as ``deformations'' of the conformal plasmas. At zero temperature, the flavor superpotential term, which drives the breaking of conformality, can be treated as a marginally irrelevant  ``deformation''  \cite{benini}, accordingly. 

At second order in $\epsilon_h$ the gravity solutions derived in \cite{d3d7plasma} provide a completely reliable description of the dual field theories in the planar limit and at strong coupling, by accounting for the backreaction of the branes supporting the flavor degrees of freedom. Thus, they allow for the calculation of the transport coefficients by means of standard procedures.

The main results are collected in the following subsection. A review of the relevant backgrounds can be found in section \ref{secbackground}, the main steps of the calculations are reported in section \ref{seccalculation} and further details are given in appendix \ref{appendix}. In section \ref{gubsec} we will provide an alternative simple calculation for the bulk viscosity, using the arguments presented in \cite{gubserbulk}. In particular we will show how an effective single-scalar five-dimensional holographic model in the Chamblin-Reall class \cite{chamblin} captures the leading conformality breaking effects due to the marginally irrelevant flavor ``deformations''. We will also show that the same approach can be successfully applied to cascading plasmas - where conformality breaking is driven by marginally relevant operators - at leading order in the perturbative expansion introduced in \cite{KTplasmapert}. We will suggest, in turn, that in the limit of small ``deformations'', there are certain universal relations for transport coefficients of gauge theory plasmas (with gravity duals) where the leading conformality breaking effects are driven by marginally (ir)relevant operators. Our proposal is a natural extension of the results in \cite{nellore}, valid for relevant or exactly marginal deformations. Finally, we will suggest possible universal relations involving the bulk viscosity and the interaction measure. As we report in the following subsection, the results of section \ref{gubsec} allow us to propose a class of universal relations for every non-conformal plasma with a dual Chamblin-Reall (effective) description.

\subsection{Main results}

Up to second order, the hydrodynamic expansion of a non-conformal plasma is known to be determined by two first order transport coefficients, \emph{i.e.} the shear and bulk viscosities, and thirteen second order transport coefficients \cite{Romatschke:2009kr}, the most important ones being the relaxation times, which are relevant for numerical hydrodynamic simulations.
We refer to \cite{Romatschke:2009kr,Buchel:2009hv} for the notation of the transport coefficients.

The bulk viscosity and a combination of the ``shear'' and ``bulk'' relaxation times $\tau_\pi, \tau_\Pi$ can be derived from the dispersion relation of the scalar hydrodynamic modes (sound channel) \cite{Baier:2007ix,Romatschke:2009kr}
\begin{equation}\label{vecdiff2}
\omega = c_s q - i \Gamma q^2 + \frac{\Gamma}{c_s}\Bigl(c_s^2\tau^{eff}-\frac{\Gamma}{2}\Bigr)q^3 + {\cal O}(q^4)\qquad {\rm where} \qquad \Gamma=\frac{\eta}{sT} \left( \frac{2}{3} + \frac{\zeta}{2\eta} \right)\,. 
\end{equation}
In this equation, $\omega$ is the frequency of the mode and $q$ its momentum; $\eta$ and $\zeta$ are respectively the shear and bulk viscosities; $s$, $T$ and $c_s$ represent the entropy density, temperature and speed of sound of the plasma. Finally, $\tau^{eff}$ is an ``effective relaxation time''\footnote{This terminology is borrowed from \cite{Buchel:2009hv}.} which for non-conformal plasmas is the combination
\begin{equation}\label{taueffgen}
\tau^{eff}=\frac{\tau_{\pi}+\frac{3\zeta}{4\eta}\tau_{\Pi}}{1+\frac{3\zeta}{4\eta}}\, .
\end{equation}

The second order coefficient $\kappa$ and a combination of $\tau_\pi$ and $\kappa^*$  can be derived from the retarded correlator of the tensorial mode \cite{Baier:2007ix,Romatschke:2009kr}
\begin{equation}\label{retcorr}
G_R^{xy,xy}=p-i \eta \omega + \Bigl( \eta \tau_\pi -\frac{\kappa}{2} +\kappa^* \Bigr)\omega^2 -\frac{\kappa}{2}q^2 + {\cal O}(q^3,\omega^3)\, ,
\end{equation}
where $p$ is the pressure.
 
In \cite{d3d7plasma} the following quantities were obtained
\begin{eqnarray}\label{old}
&& T=T_{0}\Bigl( 1-\frac18  \epsilon_h -
\frac{13}{384}\epsilon^2_h      \Bigr) \,,\qquad p=p_{0}\Bigl( 1-\frac{1}{8}\epsilon^2_h\Bigr) \,,\qquad \varepsilon-3p=\frac{p_0}{2}\epsilon_h^2\,, \rc
&&c_s=\frac{1}{\sqrt{3}}\Bigl(1-\frac{1}{12}\epsilon^2_h \Bigr) \, ,\qquad \qquad\,\, \frac{\eta}{s}=\frac{1}{4\pi}\,,
\end{eqnarray}
where $T_{0}$ and $p_{0}=\frac{\pi^5 N_c^2 T_0^4}{8 Vol(X_5)}$  are the temperature and pressure of the unflavored conformal plasmas, and $\varepsilon$ is the energy density. 

The new results in this paper, calculated up to ${\cal O}(\epsilon_h^2)$, are
\begin{eqnarray}\label{resultbulk}
\frac{\zeta}{\eta}&=&\frac19 \epsilon^2_h\,,\\ \label{resulttau}
\tau^{eff}T&=&\tau_{\pi,0}T_{0} + \frac{16-\pi^2}{128\pi}\epsilon_h^2\,,\\ \label{resultk}
\frac{T^2}{p}\kappa&=&\frac{T_{0}^2}{p_{0}}\kappa_{0}\,,\\\label{resultkstar}
\frac{T^2}{p}(\kappa^*+\eta\tau_\pi)&=&\frac{T_{0}^2}{p_{0}}\eta_{0}\tau_{\pi,0} + \frac{T_{0}^2}{p_{0}}\eta_{0}\Bigl(\frac{\tau_{\pi,0}}{8}-\frac{16+\pi^2}{128\pi T_{0}}\Bigr)\epsilon_h^2\,,
\end{eqnarray}
where 
\begin{equation}
\tau_{\pi,0}T_{0}=\frac{2-\log{2}}{2\pi}\,, \qquad \frac{T_{0}^2}{p_{0}}\kappa_{0}=\frac{1}{\pi^2}\,, \qquad \frac{T_0\eta_0}{p_0}=\frac{1}{\pi}\, ,
\end{equation}
are the corresponding values in the conformal plasmas \cite{Baier:2007ix,Bhattacharyya:2008jc}.\footnote{In the conformal case $\tau_\pi\equiv\tau_{\pi,0}=\tau^{eff}$, since $\zeta=0$.} 

Let us comment these results in turn.
The bulk viscosity (up to ${\cal O}(\epsilon_h^2)$) saturates the bound proposed by Buchel in \cite{buchelbound}, i.e. $\zeta/\eta \geq 2(1/3-c_s^2)$. In section \ref{gubsec} we obtain again this result in the framework of \cite{gubserbulk,nellore}, showing that it is just the source of the marginally irrelevant operator driving the breaking of conformal invariance (essentially the flavor superpotential term dual to the dilaton field)  the one which contributes to the bulk viscosity at leading order.
We see no reasons for the bound to be saturated at all orders.

The numerical value of $\zeta/\eta$, of order $10^{-3}$ for $\epsilon_h\sim{\cal O}(10^{-1})$ \cite{d3d7plasma}, is quite small as compared to the expected one in the near-conformal region of QCD, i.e. $\zeta/\eta\sim{\cal O}(10^{-1})$ \cite{Meyer:2007dy}.
This is expected and due to the fact that the models at hand compute just the quantum flavor effects and do not include the pure YM contribution to the trace anomaly, which is the main source to $\zeta/\eta$ in QCD.

The result for $\kappa$ is usually given in the conformal case in the combination $T_{0}\kappa_{0}/s_{0}=T_{0}^2\kappa_{0}/4p_{0}$. In the present case it reads
\begin{equation}\label{kappa}
\frac{T\kappa}{s}=\frac{T_{0}\kappa_{0}}{s_{0}}\Bigl(1-\frac{\epsilon_h^2}{8}\Bigr)\,.
\end{equation}
As such, all the results (\ref{resultbulk})-(\ref{resultkstar}) tell us that the leading corrections to the conformal values of dimensionless combinations of transport coefficients consist of a (coefficient dependent) constant times a common function of the temperature,\footnote{We thank A. Cherman, T. Cohen and A. Nellore for this observation.} which in the present case is $\epsilon_h^2 \sim \log^{-2}{T}$. This is in agreement with an extension of the results presented in \cite{nellore} to cases with marginally (ir)relevant deformations. 
In view of the general behavior discussed in section \ref{gubsec}, it would not be surprising if the numerical coefficients in (\ref{resultbulk})-(\ref{resultkstar}), written in terms of, say, $\delta_s\equiv 1-3c_s^2$ using eq. (\ref{old}), turned out to be universal in all the marginally (ir)relevant deformations of 4d conformal theories with two-derivative gravity duals.

Unfortunately, $\tau_\pi$, $\tau_\Pi$ and $\kappa^*$ cannot be disentangled in the computations reported in this paper. 
Nevertheless, the results in (\ref{resulttau}), (\ref{resultkstar}) give very interesting indications.
Comparing (\ref{taueffgen}) with (\ref{resulttau}) and (\ref{resultkstar}) using the notation\footnote{Since $\zeta/\eta\sim \epsilon_h^2$, only the zero-th order term in $\tau_{\Pi}$ is relevant in our perturbative analysis.}
\begin{equation}\label{notation}
\tau_\pi=\tau_{\pi,0}+\tau_{\pi,1}\, \epsilon_h+ \tau_{\pi,2}\,\epsilon_h^2\,, \qquad \tau_\Pi=\tau_{\Pi,0}\,, \qquad \kappa^*=\kappa^*_{0}+\kappa^*_{1}\, \epsilon_h+ \kappa^*_{2}\,\epsilon_h^2\,,
\end{equation}
where in each of these expressions it is understood that we neglect higher order terms in $\epsilon_h$, it follows that
\begin{eqnarray}  \label{tauPi}
&&\tau_{\pi,1} = \frac18 \tau_{\pi,0}\, , \qquad \tau_{\pi,2} + \frac{\tau_{\Pi,0}}{12}=\frac{17}{128}\tau_{\pi,0}+\frac{16-\pi^2}{128\pi T_{0}}\, ,\\ \label{kappastar}
&& \,\,\, \kappa^*_{0}= \kappa^*_{1}=0\, , \,\,\, \,\,\, \,\,  \tau_{\pi,2}+ \frac{\kappa^*_{2}}{\eta_{0}}=\frac{19}{384}\tau_{\pi,0}-\frac{16+\pi^2}{128\pi T_{0}}\, .
\end{eqnarray}

In section \ref{gubsec} we will note that some theories where conformal invariance is broken by marginally (ir)relevant operators are effectively described, for what concerns the bulk viscosity and speed of sound and at leading order in the deformation, by dual Chamblin-Reall models, i.e. metric plus single scalar theories with exponential potential.
For this class of models, in \cite{springer} it was found a relation between $\tau_\pi, \tau_\Pi$ which reproduces exactly the result in (\ref{tauPi}).\footnote{We thank T. Springer for pointing out this agreement.}
This means that the theories analyzed in this paper, where conformal invariance is broken by marginally (ir)relevant operators, are effectively described, at least at third order in the sound channel and at leading order in the deformation, by Chamblin-Reall models.

Crucially, the latter include also the non-conformal theory in \cite{Kanitscheider:2009as}, where the conformal transport coefficients determine all the others, in particular $\tau_\Pi=\tau_\pi$ and $\kappa^*= -\frac{\kappa}{2c_s^2}(1-3 c_s^2)$ \cite{Romatschke:2009kr}.
For the D3-D7 plasmas analyzed in this paper, postulating $\tau_\Pi=\tau_\pi$ (i.e. $\tau_{\Pi,0}=\tau_{\pi,0}$) implies from (\ref{tauPi}), (\ref{kappastar}) that $\kappa^*_2=-\kappa_0/4$, that is, using the value of the speed of sound in (\ref{old}), precisely the relation above.

Thus, ignoring the possibility of a mere coincidence, we are led to conjecture that for all the theories effectively described by a Chamblin-Reall model, the transport coefficients satisfy the relations reported in \cite{Romatschke:2009kr}:
\begin{eqnarray}\label{conjecture}
\kappa^*&=& -\frac{\kappa}{2c_s^2}(1-3 c_s^2)\,,  \qquad \tau_\pi^*=-\tau_\pi(1-3c_s^2)\,, \qquad \quad \lambda_4=0\,, \nonumber \\ 
\zeta &=& \frac{2\eta}{3}(1-3c_s^2)\,, \qquad \quad \zeta \tau_\Pi = \frac{2\eta}{3}(1-3c_s^2) \tau_\pi\,,\\ 
\xi_1&=&\frac{\lambda_1}{3}(1-3c_s^2)\,, \qquad \quad \xi_2=\frac{2\eta\tau_\pi c_s^2}{3}(1-3c_s^2)\,, \qquad  \xi_3=\frac{\lambda_3}{3}(1-3c_s^2)\,, \qquad \xi_4=0\,. \nonumber
\end{eqnarray}
In turn, the relations (\ref{conjecture}) allow to make a prediction for all the second order transport coefficients, apart from $\lambda_1, \lambda_2, \lambda_3$, up to ${\cal O}(\epsilon_h^2)$ for the D3-D7 plasmas (using the same
notation as in (\ref{notation})):
\begin{eqnarray}
\kappa^*_{0}&=&\kappa_{1}^*=0\,, \quad \kappa_{2}^*=-\frac{\kappa_{0}}{4}\,, \qquad \quad \tau_{\pi,0}^*=\tau_{\pi,1}^*=0\,, \quad \tau_{\pi,2}^*=-\frac{\tau_{\pi,0}}{6}\,, \qquad \quad \lambda_4=0\,,\nonumber \\ \tau_{\Pi,0}&=&\tau_{\pi,0}\,,\qquad \quad
\xi_{1,0}=\xi_{1,1}=0\,, \quad \xi_{1,2}=\frac{\lambda_{1,0}}{18}=\frac{\eta_0}{36\pi T_0}\,, \\ 
\xi_{2,0}&=&\xi_{2,1}=0\,, \quad \xi_{2,2}=\frac{\eta_{0}\tau_{\pi,0}}{27}\,, \qquad \quad
\xi_{3,0}=\xi_{3,1}=0\,, \quad \xi_{3,2}=\frac{\lambda_{3,0}}{18}=0\,, \qquad \quad \xi_4=0\,.  \nonumber
\end{eqnarray} 
The results in section \ref{gubsec} suggest that similar predictions can be made for the cascading plasmas.
Note that $\tau_\Pi=\tau_\pi$ would imply that $\tau^{eff}=\tau_\pi$.
Moreover, from (\ref{resulttau}) it would follow that $\tau_\pi T>\tau_{\pi,0}T_0$ at order ${\cal O}(\epsilon_h^2)$.
So, these results would support the conjecture in \cite{Buchel:2009hv} that $\tau_{\pi,0}T_0$ is a lower value for the relaxation times in theories with two-derivative\footnote{Beyond two derivatives, as for the shear viscosity, while finite coupling corrections enhance the relaxation time \cite{Buchel:2008bz}, finite $N_c$ corrections reduce it \cite{Buchel:2009nx}.} gravity duals.\footnote{Similar bounds on the speed of sound were proposed in \cite{Hohler:2009tv}, \cite{Cherman:2009tw}, and a possible universal relation among some second order coefficients were proposed in \cite{Erdmenger:2008rm}, \cite{Haack:2008xx}.}
One could also conjecture, analogously, that $\kappa_0$ constitues an upper bound for the same class of theories.

Finally, we note that the known results  \cite{Baier:2007ix,Bhattacharyya:2008jc,Kanitscheider:2009as,springer}, including the ones of this paper, are compatible with an analog of Buchel's bound of the form
\begin{equation}
\tau_{\Pi}\,\zeta\geq 2\Bigl( \frac1d -c_s^2\Bigr) \tau_{\pi}\,\eta\,,
\end{equation}
where $d$ is the number of spatial directions of the plasma ($d=3$ for the models considered in this paper). Again, analogous bounds could be proposed from the relations (\ref{conjecture}).
It is necessary to underline that these proposals are not based on first-principle arguments and they would clearly require further, more solid, confirmations.

In conclusion, the flavor effects are found to enhance both the bulk viscosity (\ref{resultbulk}) (simply because, breaking conformal invariance, they produce a non-zero $\zeta$) and the ``effective relaxation time'' (\ref{resulttau}), while 
they reduce $\kappa$ in the usual combination (\ref{kappa}) but do not modify it in the combination (\ref{resultk}).
Finally, breaking conformality, the fundamental flavors should give non-trivial coefficient $\tau_\Pi$, $\kappa^*$ (\ref{tauPi}), (\ref{kappastar}).\footnote{At least one of them is non-zero.} Unfortunately their values cannot be reliably disentangled from $\tau_\pi$ with the present gravity analysis, but we believe they satisfy the relations (\ref{conjecture}), which would imply that $\tau_\pi$ is enhanced by flavor effects.

\section{The background}\label{secbackground}
\setcounter{equation}{0}
In this section we summarize the results obtained in \cite{d3d7plasma} for the non-extremal gravity solution describing an intersection of two sets of $N_c\gg1$ D3-branes and $1\ll N_f\ll N_c$ D7-branes, where the backreaction of both stacks is taken into account. When no D7-branes are present the solution is $AdS_5 \times X_5$ with a black hole. The presence of backreacting D7-branes introduces a squashing in the internal $X_5$ (described as a $U(1)$ fibration over a four-dimensional K\"ahler-Einstein (KE) base) and the Einstein frame metric reads\footnote{With respect to \cite{d3d7plasma}'s notation we have removed the tildes from $\tilde F, \tilde S$.}
\be
ds^2 = h^{-1/2} \left[-b\,dt^2 + dx_idx_i\right] + h^{1/2}\left[S^8F^2b^{-1}dr^2 + r^2 \left(S^2 ds_{KE}^2 + F^2 (d\tau + A_{KE})^2\right)\right]\,.
\label{thebac}
\ee
Here $dA_{KE}/2=J_{KE}$ is the Kahler form of the four-dimensional base of $X_5$.
Moreover
\be
h=\frac{R^4}{r^4}\,,\qquad b=1-\frac{r_h^4}{r^4}\,,\qquad R^4\equiv \frac{Q_c}{4} = \frac{g_s\alpha'^2 N_c (2\pi)^4}{4Vol(X_5)}\,,
\label{eqh}
\ee
where the subscript $h$ stands for ``horizon'' and
\bea
F&=&1-\frac{\epsilon_*}{24}\left(1+ \frac{2r^4-r_h^4}{6r_*^4-3r_h^4}\right)+
\frac{\epsilon_*^2}{1152}\left(17-\frac{94}{9}\frac{2r^4-r_h^4}{2r_*^4-r_h^4}+
\frac59\frac{(2r^4-r_h^4)^2}{(2r_*^4-r_h^4)^2}+\right.\rc
&&\left.-\frac89 \frac{r_h^8 (r_*^4-r^4)}{(2r_*^4-r_h^4)^3}
-48 \log\frac{r}{r_*}\right)\,\,,\rc
S&=&1+\frac{\epsilon_*}{24}\left(1- \frac{2r^4-r_h^4}{6r_*^4-3r_h^4}\right)+
\frac{\epsilon_*^2}{1152}\left(9-\frac{106}{9}\frac{2r^4-r_h^4}{2r_*^4-r_h^4}+
\frac59\frac{(2r^4-r_h^4)^2}{(2r_*^4-r_h^4)^2}+\right.\rc
&&\left.-\frac89 \frac{r_h^8 (r_*^4-r^4)}{(2r_*^4-r_h^4)^3}
+48 \log\frac{r}{r_*}\right)\,\,,
\eea
up to second order in the perturbative expansion parameter $\epsilon_*\equiv\epsilon(r_*)$, where
\be
\epsilon\equiv Q_f e^{\Phi} \equiv \frac{Vol(X_3)}{16\pi Vol(X_5)}\lambda\frac{N_f}{N_c}\,,
\ee
and $\lambda\equiv 4\pi g_s N_c e^{\Phi}$; here $Vol(X_3)$ is the volume of the three-dimensional compact manifolds wrapped by the homogeneously smeared D7-branes. For the ${\cal N}=4$ plasma we have $Vol(X_5)=\pi^3$, $Vol(X_3)=2\pi^2$.  

The D7-brane sources induce a running of the dilaton $\Phi(r)$, which reads
\bea
\Phi&=& \Phi_*+\epsilon_* \log\frac{r}{r_*} + \frac{\epsilon_*^2}{72}\left[1-\frac{2r^4-r_h^4}{2r_*^4-r_h^4}
+12 \log\frac{r}{r_*} + 36 \log^2\frac{r}{r_*}+\right.\rc
&&\left.
+\frac92   
\left(Li_2\left(1-\frac{r_h^4}{r^4}\right)-Li_2\left(1-\frac{r_h^4}{r_*^4}\right)\right)
\right]\,\,.
\label{finiteTsol}
\eea
The non-trivial RR field strengths on the background are
\be
F_{(5)} = Q_c\,(1\,+\,*)\varepsilon(X_5)\,\,,\qquad
F_{(1)} = Q_f\,(d\tau + A_{KE})\,\,\Rightarrow\,\, dF_{(1)}= 2\,Q_f\,J_{KE}\,\,,
\label{f5f1}
\ee
where $\varepsilon(X_5)$ is the volume element of the internal space.

As mentioned above, at order zero in $\epsilon_*$ the solution is the standard $AdS_5 \times X_5$ black hole. The UV cutoff $\Lambda_*$, corresponding to the radial position $r_*$, is ensured to be well below the Landau pole $\Lambda_{LP}$ (mapped, in turn, to the position $r=r_{LP}$ at which the exact solution for $\Phi(r)$ blows up) if $\epsilon_*\ll 1$, and thus represents the scale up to which the solution is under control. At $\Lambda_*$ a UV completion is needed. We focus here on the IR properties of the system, such that we can safely discard contributions coming from the UV completion, which are power-like terms suppressed as $(\Lambda_{IR}/\Lambda_{*})^n$.
Thus, in the computations below we systematically discard the (subleading) corrections in $r_h/r_*$. In this regime the previous expressions for $F$, $S$ and $\Phi$ get the simpler form
\bea
F &=& 1 - \frac{\epsilon_h}{24} + \frac{17}{1152}\epsilon_h^2 -\frac{\epsilon_h^2}{24}\log\frac{r}{r_h} \equiv F_h -\frac{\epsilon_h^2}{24}\log\frac{r}{r_h}\,,\rc
S &=& 1 + \frac{\epsilon_h}{24} + \frac{1}{128}\epsilon_h^2 + \frac{\epsilon_h^2}{24}\log\frac{r}{r_h}\equiv S_h  + \frac{\epsilon_h^2}{24}\log\frac{r}{r_h}\,,\rc
\Phi &=& \Phi_h + \epsilon_h \log\frac{r}{r_h} + \frac{\epsilon_h^2}{6} \log\frac{r}{r_h} + \frac{\epsilon_h^2}{2} \log^2\frac{r}{r_h} + \frac{\epsilon_h^2}{16} Li_2\left(1-\frac{r_h^4}{r^4}\right)\,.
\label{simple}
\eea
Notice that, being the temperature $T$ proportional to $r_h$ at leading order \cite{d3d7plasma}, one has
\be\label{epsilonrun}
\epsilon_h = \epsilon_* + \epsilon_*^2 \log\frac{r_h}{r_*}\,,\qquad T\frac{d\epsilon_h}{dT} = \epsilon_h^2\,,
\ee
which implies that while $\epsilon$ runs with the energy scale (and thus with the temperature), $\epsilon^2$ is constant if we neglect $\epsilon^3$ and higher order terms.
The interaction measure $(\varepsilon-3p)/T^4$ and the shifted speed of sound $c_s^2- 1/3$ (see eq. (\ref{old})), which are zero for conformal plasmas, scale like $\epsilon_h^2$ as an effect of quantum conformality breaking due to the dynamical massless flavors  \cite{d3d7plasma}. Those observables are thus independent on the temperature at next-to-leading order in $\epsilon_h$. We will come back to this observation in section \ref{gubsec}.


\subsection{Reduced action}
In the following we will study hydrodynamic transport properties of the plasmas described above. We first rewrite our ten-dimensional model as an effective five-dimensional one by integrating over the internal manifold $X_5$.
The five-dimensional reduction of the action reads \cite{benini}
\begin{equation}\label{5daction}
S_5 = \frac{Vol(X_5)}{2\kappa_{10}^2} \int d^5 x \sqrt{-\det g} \left[ R[g] - \frac{40}{3}(\partial f)^2 - 20 (\partial w)^2 - \frac{1}{2}(\partial\Phi)^2 - V(\Phi,f,w) \right]\, ,
\end{equation}
where $g_{mn}$ is the five-dimensional metric, $R[g]$ is its scalar of curvature and the potential describing the interactions of the three scalars  reads
\begin{equation}
V(\Phi,f,w) = 4 e^{\frac{16}{3}f+2w} \left( e^{10w}-6 + Q_f e^{\Phi}\right) + \frac{1}{2} Q_f^2 e^{\frac{16}{3} f -8w +2\Phi} + \frac{Q_c^2}{2}e^{\frac{40}{3} f}\, .
\label{potV}
\end{equation}
Hereafter we set $R=1$ and $\alpha'=1$ for simplicity. In these units $Q_c=4$ as can be read from (\ref{eqh}).

To get the above results one starts from a reduction ansatz of the form
\be
ds_{10}^2 =  e^{\frac{10}{3}f}g_{mn} dx^m dx^n + e^{-2(f+w)}ds_{KE}^2 + e^{2(4w-f)} (d\tau+A_{KE})^2\,.
\ee
On the background (\ref{thebac}) we have
\bea
f = -\frac15\log\left( S^4F \right)\, , \qquad w = \frac15 \log\left( F/S \right)\,  .
\label{BCdef}
\eea
The five-dimensional non-extremal background metric reads
\be
ds_5^2 = r^2 e^{-\frac{10}{3}f} [- b\,dt^2 + dx_i dx_i] + e^{-\frac{40}{3}f} b^{-1} \frac{dr^2}{r^2}\equiv -c_T^2 dt^2 + c_X^2dx_i dx_i + c_R^2dr^2\,.
\label{5dbacmetric}
\ee

From the dimensional reduction we see that we have three scalar fields. In the perturbative expansion in $\epsilon_*$ our models (both in the extremal and in the non-extremal case) can be seen as ``deformations'' of the unflavored $AdS_5\times X_5$ (BH) solutions. The $AdS$ background  is a minimum of the potential at order zero in $\epsilon_*$, in which case $f=0$ and $w=0$. The field $f$ is dual to an irrelevant operator of dimension $\Delta=8$ whose form is $\tr F^4$. It drives a deformation from the $AdS_5\times X_5$ to the non-near horizon D3-brane background. The field $w$ is dual to a vev for an irrelevant operator of dimension $\Delta=6$. It is of the form $\tr ({\cal W}_{\alpha}{\cal W}^{\alpha})^2$ and is the responsible of the squashing of the transverse four-dimensional Kahler-Einstein base and the fibration. The dilaton $\Phi$ is dual to the insertion of a marginally irrelevant operator, actually the flavor term in the field theory ($T=0$) superpotential \cite{benini}. This is the source term which is responsible for the breaking of conformal invariance at the quantum level.

Notice that the action (\ref{5daction}), and so the equations of motion for some of the perturbations, coincides with the one in \cite{benincasa} (but for the definition of the potential). The  equations of motion for the fluctuations not coupling to the scalar fields (that is, insensitive to the potential) can be read in \cite{mas}. 

\section{Fluctuations}\label{seccalculation}
\setcounter{equation}{0}
Following the standard procedure \cite{Kovtun:2005ev}, we will consider fluctuations of the fields present in the five-dimensional action. It is easily seen from the expansion of the DBI action that the perturbations considered in the following do not couple to those of the flavor branes at the linearized level, thus the relevant set of fluctuations we need to consider is
$
\Psi (r)  \to  \Psi(r) + \delta \Psi(x^\mu,r)
$,
with $\Psi=\left\{ g_{mn}, \Phi, f, w \right\}$.

We will assume that the perturbations take a planar wave form in Minkowski space,
$
\delta\Psi(x^\mu,r) = e^{-i(\omega t - q z)} \Psi(r) 
$.
Thus, these fields can be classified according to their transformation under the little group $SO(2)$, which is the remaining symmetry of the system (rotations in the $x-y$ plane). We define
\begin{eqnarray}
\delta g_{tt}(x^\mu,r) &=& e^{-i(\omega t - q z)} c_T^2(r) \ch_{tt}(r)\, , \\
\delta g_{mn}(x^\mu,r) &=& e^{-i(\omega t - q z)} c_X^2(r) \ch_{mn}(r)\, , \quad (m,n) \neq (t,t)\\
\delta \Phi(x^\mu,r) &=& e^{-i(\omega t - q z)} \cph(r)\, , \\
\delta f(x^\mu,r) &=& e^{-i(\omega t - q z)} \cb(r)\,, \\
\delta w(x^\mu,r) &=& e^{-i(\omega t - q z)} \cc(r)\, ,
\end{eqnarray}
choosing the gauge $\ch_{rm}(r)=0$. The classification of the different channels gives a tensorial mode ($\ch_{xy}$), vectorial modes ($\ch_{tx}$, $\ch_{zx}$, $\ch_{ty}$, $\ch_{zy}$) and scalar modes ($\ch_{tt}$, $\ch_\pperp \equiv \ch_{xx} + \ch_{yy}$, $\ch_{zz}$, $\ch_{tz}$, $\cph$, $\cb$, $\cc$). 
Each kind of perturbation can be expressed in term of gauge invariant quantities under the residual gauge symmetry  \cite{Kovtun:2005ev,benincasa,mas}
\begin{eqnarray*}
\mathrm{Tensorial} & \to & \cz_{T} = \ch_{xy} \, , \\
\mathrm{Vectorial} & \to & \cz_{V} = q \ch_{tx} + \omega \ch_{zx} \, , \\
\mathrm{Scalar} & \to & \cz_{S} = 2 \ch_{zz} + 4 \frac{q}{\omega} \ch_{tz} - \left[ 1- \frac{q^2}{\omega^2} \frac{c_T' c_T}{c_X' c_X} \right] \ch_\pperp + 2 \frac{q^2}{\omega^2} \frac{c_T^2}{c_X^2} \ch_{tt} \, ,\nonumber\\
& & \cz_{\cph} = \cph - \frac{\Phi' }{\log'\left[ c_X^4 \right]} \ch_\pperp  \, , \\
& & \cz_{\cb} = \cb - \frac{f'}{\log'\left[ c_X^4 \right]} \ch_\pperp \, , \\
& & \cz_{\cc} = \cc - \frac{w'  }{\log'\left[ c_X^4 \right]} \ch_\pperp \, . \nonumber
\end{eqnarray*}
In all the cases, studying the differential equations at the horizon we find that the solutions behave as $(\frac{r}{r_h}-1)^{\pm i \frac{\wn T_{0}}{2T}}$, where $\wn = \omega / (2 r_h)$ and $r_h=\pi T_0$.  We choose the index with negative sign to have incoming boundary conditions at the black hole horizon. Moreover, as in \cite{screeningmesons}, we impose that the fluctuations vanish at the UV cutoff scale related to $r_*$: crucially, our results will turn out to be independent of $r_*$ up to suppressed terms in powers of $r_h/r_*$. 

In the following we are going to study the tensorial and scalar perturbations, while the study of the vectorial perturbation is relegated to appendix \ref{appendix}.

\subsection{Tensorial perturbation}

 We scale $\wn \rightarrow \lambda_{hyd} \wn,\,  \qn \rightarrow \lambda_{hyd} \qn$, where $\qn=q/(2r_h)$ and $\lambda_{hyd}$ is a parameter keeping track of the order of the hydrodynamic expansion. Define
 \be\label{ansatztensorial}
 \cz_T = C_T \left(1-\frac{r_h^4}{r^4}\right)^{-i \frac{\wn T_{0}}{2 T}} \sum_{j=0}^2\sum_{k=0}^2 \cz_T^{j,k}\, \lambda_{hyd}^j\, \epsilon_*^k\,,
 \ee
where higher order terms in $\epsilon_*$ and $\lambda_{hyd}$, which we will not study, are not taken into account. The equation satisfied by the perturbation is reported in appendix \ref{appendix}.
As expected, Dirichlet conditions cannot be imposed, showing the absence of a dispersion relation in this channel \cite{Kovtun:2005ev}. However one can write the hydrodynamic expansion of the retarded correlator. To do this we have to evaluate the action on-shell. This action is singular when evaluated at $r=r_*\to\infty$ (it goes as $r_*^4$). To cure this divergence we have to add the following counterterms \cite{counterterms} \begin{equation}\label{totalaction}
S_{bulk} \to S_{bulk} + \frac{Vol(X_5)}{2\kappa_{10}^2} \int d^4 \xi \,2\sqrt{-\gamma} \,K + \frac{Vol(X_5)}{2\kappa_{10}^2} \int d^4 \xi \sqrt{-\gamma}\, \Bigl({\cal W[\phi]} - \frac{1}{2} C[\phi]R[\gamma] \Bigr)\,  ,
\end{equation}
where $K$ is the scalar associated to the extrinsic curvature, $\gamma$ is the four-dimensional metric at the boundary, $C[\phi]$ is a function of the scalars and $\cal W[\phi]$ is the superpotential 
\begin{equation}
{\cal W}[f,w,\Phi] = e^{\frac{5}{3}f} \left[ Q_c\, e^{5f} + Q_f e^{f-4w+\Phi} - 4 e^{f+6w} - 6 e^{f-4w} \right] \, ,
\end{equation}
from which the potential (\ref{potV}) can be derived as
\begin{equation}
V = \frac{1}{2} \left[  \frac{3}{80} \left(\frac{\partial {\cal W} }{\partial f}\right)^2 + \frac{1}{40} \left(\frac{\partial {\cal W} }{\partial w}\right)^2 +  \left(\frac{\partial {\cal W} }{\partial \Phi}\right)^2 \right] - \frac{1}{3} {\cal W} ^2 \, .
\end{equation}

The function $C[\phi]$ satisfies the differential equation \cite{counterterms}
\begin{equation}
\frac{1}{2} - \frac{1}{4} \left[  \frac{3}{80} \frac{\partial {\cal W} }{\partial f}\frac{\partial {C} }{\partial f} + \frac{1}{40}\frac{\partial {\cal W} }{\partial w}\frac{\partial {C} }{\partial w} +  \frac{\partial {\cal W} }{\partial \Phi}\frac{\partial {C} }{\partial \Phi} \right] + \frac{1}{12}C\,{\cal W} = 0 \,\, .
\end{equation}
Although we do not know the exact form of $C[\phi]$, we can extract physical results, since the divergence balanced by this function goes in the UV as $r_*^2$, being the next-to-leading order suppressed as $r_*^{-2}$, \emph{i.e.}, it does not affect the finite part from which the hydrodynamic transport coefficients are obtained. For completeness we give its leading behavior, needed to cancel the divergence
\begin{equation}
C[f,w,\Phi] \approx 1 + \frac{23}{108}\epsilon_* - \frac{371}{23328} \epsilon_*^2 +{\cal O}\left( r^{-4}\right)\, \, .
\end{equation}

The Fourier transformed, quadratic-in-fluctuations, on-shell boundary action is
\begin{equation}
S = \frac{Vol(X_5)}{2\kappa_{10}^2} \int d^4 k H_{-k}{\cal F}(k,r_*) H_k \, ,
\end{equation}
with $H_k$ the boundary value of the fluctuation.  The retarded correlator of the corresponding components of the energy momentum tensor
\begin{equation}
G_R^{xy,xy}(\omega,\,q) = - i \int dt d^3x e^{i(\omega t-qz)}\Theta(t) \langle [T_{xy}(t,\vec x), T_{xy}(0,\vec 0)]  \rangle \, ,
\end{equation}
is related to the on-shell action by \cite{Son:2002sd}
\begin{equation}
G_R^{xy,xy}(\omega,\,q) = -2 \,{\rm Im}[{\cal F}(k, r_{*})]\, .
\end{equation}

Plugging the solution of the equation of motion for $\cz_T$, equation (\ref{eqfortensor}), in the finite action (\ref{totalaction}), it is straightforward to derive the flux ${\cal F}(k,r_*)$,
and so the correlator $G_R^{xy,xy}(\omega,\,q)$.
Using also (\ref{epsilonrun}), we get
\begin{eqnarray}
G_R^{xy,xy}&=& \frac{\pi^5 N_c^2 T^4_{0}}{8 Vol(X_5)} \Bigl( [1-2 i \wn -2 \qn^2 + 2\wn^2(1-\log{2})]-\frac{ i \wn +2 \qn^2 - 2\wn^2(1-\log{2})}{4}\epsilon_h+\nonumber\\
&& - \frac{24+19 i \wn -4 \qn^2 + 2\wn^2(2+3\pi^2+22\log{2})}{192}\epsilon_h^2 \Bigr)\,,
\end{eqnarray}
which, compared to (\ref{retcorr}) and using the expression for the temperature in (\ref{old}), confirms the values of the pressure $p$ and shear viscosity $\eta$ from \cite{d3d7plasma} and gives the new results in (\ref{resultk}), (\ref{resultkstar}).

\subsection{Scalar perturbations}
Focusing now on the scalar fluctuations we write for each perturbabion $\cz_{A=S,\cb,\cc,\varphi}$ the ansatz
\begin{eqnarray}
\wn & = & \sum_{k=0}^2 c_{s,k}\,\epsilon_*^k \,\qn- 2\, i\,\sum_{k=0}^2 \gamma_{k}\,\epsilon_*^k \,  \qn^2  + 4\,  \sum_{k=0}^2 t_{k}\,\epsilon_*^k \, \qn^3\,,\\ 
 \cz_A & = & C_A \left(1-\frac{r_h^4}{r^{4}}\right)^{-i\frac{\wn T_0}{2T}} \sum_{j=0}^2\sum_{k=0}^2 \cz_A^{j,k}\, \qn^j\, \epsilon_*^k\,, 
\end{eqnarray}
where $\gamma_k, t_k$ are the order $\epsilon_*^k$ coefficients of the adimensional combinations 
\be
\gamma\equiv \pi T_0\, \Gamma\,,\qquad t\equiv (\pi T_0)^2\,\frac{\Gamma}{c_s}\Bigl(c_s^2\tau^{eff}-\frac{\Gamma}{2}\Bigr)\,.
\ee
Higher order terms in $\epsilon_*$ and $\qn$, which we will not study, are not considered. 

The relevant equations for the perturbations are reported in appendix \ref{appendix}.
The calculation can be performed imposing regularity at the horizon. Once this is obtained, one can ask for Dirichlet conditions at the UV cutoff, $r_*$, eventually taking the limit $r_*\to\infty$. Here we present only the results relevant for the physical observables. The solution is given in appendix  \ref{appendix}. 
With the Dirichlet condition at the boundary we find
\begin{eqnarray}\label{res}
c_{s,0} & = & \frac{1}{\sqrt{3}} \, , \qquad \qquad c_{s,1} = 0 \, , \qquad \qquad c_{s,2} = -\frac{1}{12\sqrt{3}} \, , \\
 \gamma_0 & = & \frac{1}{6} \, , \qquad \qquad \quad \gamma_1 = \frac{1}{48} \, , \qquad \qquad \gamma_2 = \frac{17-16\log[\frac{r_*}{r_h}]}{768} \, ,\nonumber \\
 t_0  & = & \frac{3-2\log{2}}{24\sqrt{3}} \, , \quad t_1   = \frac{3-2\log{2}}{96\sqrt{3}} \, , \quad t_2 = \frac{57 - 3 \pi^2 - 22 \log{2} - 24 (3 - 2\log{2}) \log[\frac{r_*}{r_h}]}{2304 \sqrt{3}} \, ,\nonumber
 \end{eqnarray}
 which confirms the result for the speed of sound found with the thermodynamics in \cite{d3d7plasma} and, using (\ref{vecdiff2}), (\ref{old}) and (\ref{epsilonrun}), gives the new results reported in (\ref{resultbulk}), (\ref{resulttau}).
 
\section{A simple perturbative approach to first order hydrodynamics}
\setcounter{equation}{0}
\label{gubsec}
Let us consider a simple five-dimensional gravity model with a minimally coupled scalar field $\phi$
\be
S_5 = \frac{1}{2\kappa_5^2}\int d^5x \sqrt{-\det g}\left[R[g] - \frac{1}{2}(\partial\phi)^2- V(\phi)\right]\,,
\ee
and assume that this model admits an $AdS$ (black hole) vacuum - dual to a four-dimensional (thermal) CFT - when the scalar field is turned off. Considering thermal cases in which $\phi$ is dual to a source for a relevant operator of asymptotic dimension $2<\Delta<4$, the authors of \cite{gubserbulk} provided a simple expression relating the bulk viscosity\footnote{This is the only non trivial first order hydrodynamic coefficient, since $\eta/s=1/(4\pi)$ for any strongly coupled plasma with a two-derivative gravity dual \cite{kss}.} of the dual field theory plasma with the five-dimensional scalar potential:
\be
\frac{\zeta}{\eta}= |h^{(0)}_{11}(\phi_h)|^2\,\left(\frac{V'(\phi_h)}{V(\phi_h)}\right)^2\,.
\label{zetagub}
\ee
An analogous approximate expression for $c_s^2-1/3$ was proposed in \cite{gubserspeed}. In (\ref{zetagub}) the prime means derivative w.r.t. $\phi$ and $\phi_h$ is the value of the field at the horizon. The coefficient $h^{(0)}_{11}(\phi_h)$ is determined by solving the equation of motion for the $SO(3)$ invariant fluctuation $H_{11}=H_{22}=H_{33}\equiv e^{i\omega t}h_{11}(\phi)$ of the three-dimensional spatial metric components\footnote{In fact the bulk viscosity is related, via Kubo formulas, precisely to the (low frequency limit of the) $SO(3)$ invariant retarded Green's function of the operator $T_{11}+T_{22}+T_{33}$, where $T_{\mu\nu}$ is the field theory stress-energy tensor.} at $\omega=0$. In the $r=\phi$ gauge, in which the five-dimensional radial coordinate is identified with the background scalar field $\phi$, the equation is given by
\be
h_{11}''= \left[-\frac{1}{3A'} - 4A' + 3Y' -\frac{b'}{b}\right]h_{11}' + \frac{b'}{b}\left[\frac{1}{6A'} -Y'\right] h_{11}\,,
\label{gubeq}
\ee
where the functions $A,Y$ enter the background metric as
\be
ds^2 = e^{2A}[-b\,dt^2+dx_idx_i] + e^{2Y}\frac{d\phi^2}{b}\,.
\label{5dgubmet}
\ee
In order to solve eq. (\ref{gubeq}) one imposes regularity at the horizon (selecting only incoming waves in the $\omega\neq0$ case) and the (normalized to fix the residual three-dimensional scale factor) boundary condition $h_{11}\rightarrow1$ at the asymptotic $AdS$ boundary. 

In \cite{nellore} (see also \cite{yarom}) it was shown that the coefficient $h_{11}^{(0)}(\phi_h)$ depends only on $\Delta$ (and not on the details of the five-dimensional potential) at leading order in a perturbative expansion around the conformal background (i.e. in a large $T$ limit if the deformation related to $\phi$ is relevant). The general results of \cite{nellore,yarom} can be eventually extrapolated to exactly marginal deformations, in which case they consistently give trivial hydrodynamic coefficients (i.e. $\zeta=c_s^2-1/3=0$); however, they do not  apply to theories where conformality breaking is driven by marginally (ir)relevant operators, like, notably, Yang-Mills as well as the D3-D7 and the cascading plasmas \cite{KTplasmapert}. In \cite{gursoy} it was shown that, at $T\gg T_c$, $h_{11}^{(0)}(\phi_h)\rightarrow1$ in a five-dimensional phenomenological holographic dual to Yang-Mills. In the following we will show that the same result applies to the D3-D7 and the cascading plasmas in analogous asymptotic regimes.

It is relevant to notice that, as it was shown in \cite{gubserbulk}, eq. (\ref{gubeq}) always admits a constant solution (which in turns implies that $h_{11}^{(0)}(\phi_h)=1$ to satisfy the boundary conditions) for Chamblin-Reall backgrounds \cite{chamblin}, where $V=V_0\,e^{\gamma\phi}$, with $V_0<0$ and $\gamma$ constants. As we will see in the following, the 5d duals to the D3-D7 plasmas effectively behave like the Chamblin-Reall models, at leading order in the $\epsilon$-expansion around the conformal fixed point. As we will show, the same applies to the cascading plasma \cite{KTplasmapert} in an analogous ``perturbative'' regime. It is worth underlining that other relevant non-conformal plasmas \cite{mas,Kanitscheider:2009as} have a dual gravity description precisely given by a model in the Chamblin-Reall class \cite{gubserspeed,gubserbulk}. We argue that this is the reason why the above mentioned systems (at leading ``perturbative'' order, in the case of the D3-D7 and the cascading plasmas) have so many common features (for example they all saturate the bulk viscosity bound proposed in \cite{buchelbound}) despite having different microscopical content. 

\subsection{Bulk viscosity of D3-D7 plasmas}
Reducing the D3-D7 models to five dimensions (see eqs. (\ref{5daction}), (\ref{potV})) we have seen that there are three scalar fields in the action. However, only one of them, namely the dilaton, is dual to the source for the (marginally irrelevant) deformation driving our theories away from conformality \cite{benini}.  It is thus conceivable that the bulk viscosity, which is turned on when conformality is broken, is primarily determined by the dilaton field in an expansion around the $AdS$-BH solutions. 

To better understand the role played by the various scalars, let us consider the quantity
\be
{\cal V}_\phi\equiv\left(\frac{V'(\phi_h)}{V(\phi_h)}\right)^2\,,
\label{calV}
\ee
in three different cases, where we identify $\phi$ with one of the three fields entering in the potential (\ref{potV}), taking the other two scalars fixed to their background values. At order $\epsilon_h^2$ we get
\begin{equation}
{\cal V}_{f,w}= 0 \, , ~~~ {\cal V}_{\Phi}=\frac{\epsilon_h^2}{9}\, .
\end{equation}
This indicates that the bulk viscosity (and the speed of sound) can be determined in our models by considering {\it just the dilaton $\Phi$} as the ``active'' field in the game. The other two scalars, $f$ and $w$, do not contribute at leading order and they can be fixed to their background values. Due to this observation we can immediately apply the recipes of \cite{gubserbulk}, based on a single-scalar five-dimensional model, to our cases. 

Let us define, at first order in $\epsilon_h$, a new radial variable
\be
\phi\equiv\Phi-\Phi_h = \epsilon_h \log\frac{r}{r_h}\,  \Rightarrow \, \frac{r}{r_h}=e^{\frac{\phi}{\epsilon_h}}\,,
\ee
from which we can re-express our five-dimensional metric (\ref{5dbacmetric}), in the form (\ref{5dgubmet}) with 
\be
A= \frac{\phi}{\epsilon_h} + \frac{\epsilon_h}{24}(1+\phi)+ {\rm const}\,,\quad
Y = \frac{\epsilon_h}{6}(1+\phi) + {\rm const}\,.
\ee
From these expressions we find that the term in $h_{11}$ in (\ref{gubeq}) vanishes (up to ${\cal O}(\epsilon_h^3)$ terms). Just as for the Chamblin-Reall models \cite{gubserbulk}, $h_{11}={\rm constant}$ is thus a solution to (\ref{gubeq}) at leading order. This implies that $h_{11}^{(0)}(\phi_h)=1$ to satisfy the boundary conditions. As a result, Buchel's bound on the bulk viscosity \cite{buchelbound} is saturated. In fact we obtain 
\be
\frac{\zeta}{\eta} = \frac{V'(\phi_h)^2}{V(\phi_h)^2} = \frac{\epsilon_h^2}{9}\,.
\ee
Notice in turn that $c_s^2-1/3 = -\epsilon_h^2/18$ (as computed in \cite{d3d7plasma} and in the previous section) can be expressed as
\be
c_s^2 - \frac13 = -\frac{1}{2} \frac{V'(\phi_h)^2}{V(\phi_h)^2}\,,
\ee
just as it happens in the Chamblin-Reall cases \cite{gubserspeed}.

\subsection{The bulk viscosity of the cascading plasma}
Many relevant five-dimensional gravity models, dual to well studied non-conformal field theories, contain more than one scalar.  A well known example is given by the cascading conifold theory \cite{kt,ks} whose hydrodynamics has been studied in great detail by Buchel and collaborators (see for example \cite{buchelcasca}). This theory describes the low energy dynamics of $N$ regular and $M$ fractional D3-branes on the conifold. When $M=0$, the theory has gauge group $SU(N)\times SU(N)$ and it is conformal with an $AdS_5\times T^{1,1}$ dual \cite{kw}.  The addition of fractional branes modifies the gauge group to $SU(N+M)\times SU(N)$ and breaks conformal invariance. The combination $g_1^{-2}-g_2^{-2}$ of the gauge couplings, in fact, acquires a logarithmic running with the scale $g_1^{-2}-g_2^{-2}\sim M\log(\mu/\Lambda_{IR})$. The marginally relevant operator $Tr F_1^2 - Tr F_2^2$ is mapped to a massless scalar field  in the dual five-dimensional gravity description (see e.g. \cite{milano} for a complete scalar/operator map in the deformed conifold theory). This is actually the supergravity modulus arising from the integral of $B_2$ over the two-cycle of the conifold. The other massless field, the dilaton, is dual to $Tr F_1^2 + Tr F_2^2$ and (differently from the D3-D7 cases examined above) is just a constant on the background at $T=0$. The five-dimensional effective action reads
\bea\label{5dactionKT}
&&S_5=\frac{Vol(T^{1,1})}{2\kappa_{10}^2}\int d^5 x \sqrt{-\det g} \left[ R[g] - {\cal L}_{kin}-V(f,w,\Phi,K) \right]\,,\rc
&&{\cal L}_{kin}=\frac{40}{3}(\partial f)^2 + 20 (\partial w)^2 + \frac{1}{2}(\partial\Phi)^2 +\frac{1}{4P^2} e^{-\Phi+4f+4w} (\partial K)^2\,,\rc
&&V(f,w,\Phi,K) = 4 e^{\frac{16}{3}f+2w} \left( e^{10w}-6 \right) + P^2 e^{\Phi+\frac{28}{3}f-4w} + \frac12 K^2 e^{\frac{40}{3}f}\,.
\eea
In $\alpha'=1$ units, $P\sim g_s M$ and $K$ is proportional to the effective number of regular D3-branes, which is running with the scale if $M\neq0$ \cite{kt}.  At $P=0$ the previous action has an $AdS$ (BH) vacuum where (setting the $AdS$ radius to one) $K=K_*=4$. On this background $\Phi=f=w=0$. The fields $f,w$ are mapped to irrelevant operators with $\Delta=8,6$ just as in the D3-D7 setup.

A regular non-extremal solution of the equations of motion following from (\ref{5dactionKT}), has been found using a perturbative approach \cite{KTplasmapert} analogous to the one we have adopted in the D3-D7 case, namely by means of an expansion in $\delta=P^2/K_*$.  This parameter has a non zero beta function at ${\cal O}(\delta^2)$: its logarithmic running with the temperature (from which it follows that $\delta\ll1$ at large $T\gg\Lambda_{IR}$ \cite{KTplasmapert}) can be thus neglected at ${\cal O}(\delta)$. At this order and in our units, $\delta=P^2/4$. If we rewrite $K= 4 + \sqrt{2}P\chi$ we can see that the field $\chi$ has canonically normalized kinetic term and it is massless, around the $AdS$ (BH) vacuum. This is precisely the field dual to the $Tr F_1^2 - Tr F_2^2$ operator mentioned above.

In the non-extremal case, the functions $f,w,\Phi$ (resp. $\chi$) receive the first corrections at ${\cal O}(\delta)$ (resp. ${\cal O}(\sqrt{\delta})$). If we now consider the quantity ${\cal V}_\phi$ defined in (\ref{calV}) we get 
\begin{equation}
{\cal V}_{f,w,\Phi}= 0 \, , ~~~ {\cal V}_{\chi}=\frac{8}{9}\delta\, ,
\end{equation}
at leading order. Precisely as in the D3-D7 case, the only ``active'' scalar, for what concerns the leading perturbative contribution to the bulk viscosity, is the one related to the marginal (in this case marginally relevant) deformation which sources the breaking of conformal invariance. It is possible to verify, as above, that $h_{11}^{(0)}(\phi_h)=1$ at leading order, such that
\be
\frac{\zeta}{\eta} = \frac{V'(\phi_h)^2}{V(\phi_h)^2}= \frac{8}{9}\delta\,,
\ee
nicely reproducing the results found in \cite{buchelcasca} by means of the alternative analysis of the hydrodynamical pole in the stress-energy tensor. Notice, again, that the speed of sound $c_s^2-1/3= -4\delta/9$ (as computed in \cite{buchelcasca}) satisfies the relation
\be
c_s^2 - \frac13 = -\frac{1}{2} \frac{V'(\phi_h)^2}{V(\phi_h)^2}\,,
\ee
just as it happens for the Chamblin-Reall models. From the above expressions we see that Buchel's bound is saturated at leading order, consistently with the results found in \cite{buchelcasca}.

\subsection{Comments}

From the above results it is tempting to propose that
\be
\frac{\zeta}{\eta} = \frac{V'(\phi_h)^2}{V(\phi_h)^2} = 2\Bigl(\frac13-c_s^2\Bigr)\,,
\label{argue}
\ee
in the vicinity of a conformal fixed point - i.e. at leading order in a ``small'' (resp. ``large'') $T$ ``perturbative'' expansion - for every four-dimensional plasma with a five-dimensional two-derivative gravity dual where conformality breaking is driven at the quantum level by a marginally irrelevant (resp. marginally relevant) operator. A common feature of this kind of plasmas is that the perturbative expansion parameter $\delta$ runs as the inverse of the logarithm of the temperature and hence $\beta_T[\delta]\equiv T(\partial\delta/\partial T)\sim \pm \delta^2$. This in turn implies that $\beta_T[(\varepsilon -3p)/T^4]$ is subleading w.r.t. $(\varepsilon-3p)/T^4$: the latter (as well as $c_s^2-1/3$) is thus effectively constant at leading order.

The relations (\ref{argue}), easily extended to $d\neq3$ space dimensions (just replacing $1/3$ with $1/d$ on the r.h.s), are precisely satisfied by other known non-conformal models \cite{mas,Kanitscheider:2009as} whose dual gravity description is in the Chamblin-Reall class. The results of \cite{nellore,yarom}, instead, show that gravity duals of relevant deformations do not belong to this class, as eqns. (\ref{argue}) do not hold. 

A well studied example of a plasma where conformality breaking is driven by relevant deformations is the ${\cal N}=2^*$ one \cite{bucheln2}. In this case the conjectured bound in \cite{buchelbound} is not saturated, the perturbative expansion parameters are of the form
\be
\delta =  \left(\frac{m}{T}\right)^{(4-\Delta)}\,,\quad {\rm so\,\,that}\quad  \beta_T [\delta] = (\Delta-4)\delta\,,
\ee
and $(\varepsilon-3p)/T^4 \sim \delta^2$. This means that the interaction measure (as well as $c_s^2-1/3$) is not a constant at leading order: $\beta_T[(\varepsilon -3p)/T^4] \sim \delta^2$ is of the same order as $(\varepsilon-3p)/T^4$. This in turn implies that the model does not have an effective Chamblin-Reall dual description.

The above examples suggest that Buchel's bound could be perturbatively saturated whenever $\beta_T[(\varepsilon-3p)/T^4]$ is subleading, i.e. that, more generically
\be
\frac{\zeta}{\eta} - 2\left(\frac13-c_s^2\right) \sim \frac{f(\delta,\lambda,T)}{sT} \left[T\frac{\partial}{\partial T}-4\right](\varepsilon-3p)\,,
\label{fguess}
\ee
for some model-dependent dimensionless function $f(\delta,\lambda,T)$.\footnote{Using the results in \cite{yarom} we find that $f\rightarrow (2/9) (\Delta-2)/(\Delta-4) - (\pi/9)\cot(\pi\Delta/4)$ at large $T$ for strictly relevant deformations of planar strongly coupled gauge theories with a gravity dual.} This proposal, modulo a slight modification needed to accommodate the results collected above, is analogous to those presented in \cite{lowen} by means of exact sum rules and a certain assumption for the spectral density of the trace of the stress energy tensor. Notice that our formula (\ref{fguess}) is meant to apply to four-dimensional strongly coupled plasmas with massless flavors having a two-derivative five-dimensional gravity dual, in a regime where any possible ``deconfining temperature'' $T_c$ can be neglected.\footnote{In our D3-D7 models $T_c=0$ exactly, while $T_c\ll T$ in the perturbative regime for the cascading plasma.}

\vskip 15pt
\centerline{\bf Acknowledgments}
\vskip 10pt
\noindent
We are grateful to Aleksey Cherman, Thomas Cohen, Javier Mas, Abhinav Nellore, Angel Paredes, Alfonso V. Ramallo and Todd Springer for relevant observations. F. B. is supported  by the Belgian Fonds de la Recherche Fondamentale Collective (grant 2.4655.07), by the Belgian Institut Interuniversitaire des Sciences Nucl\'eaires (grant 4.4505.86) and the Interuniversity Attraction Poles Programme (Belgian Science Policy). A. C. is supported by the FWO -Vlaanderen, project G.0235.05 and by the Federal Office for Scientific, Technical and Cultural Affairs through the Interuniversity Attraction Poles Programme (Belgian Science Policy) P6/11-P.
J.T. is supported by the MEC and  FEDER (grant FPA2008-01838), the Spanish Consolider-Ingenio 2010 Programme CPAN (CSD2007-00042), Xunta de Galicia (Conselleria de Educacion, grant PGIDIT06PXIB206185PR, project number INCITE09 206 121 PR) and by MEC of Spain under a grant of the FPU program. J. T. would like to thank the Perimeter Institute for hospitality at early stages of this paper.

{ \it F. B. and A. L. C. would like to thank the Italian students, parents, teachers and scientists for
their activity in support of public education and research.}
\appendix

\section{Details of the calculations}\label{appendix}
\setcounter{equation}{0}

\subsection*{Tensorial perturbations}

From the action (\ref{5daction}) it follows that the equation satisfied by the tensorial perturbation $\cz_{T} = \ch_{xy}$ is
\begin{equation}\label{eqfortensor}
\cz_{T}''+\log'{\Bigl(\frac{c_T c_X^3}{c_R}\Bigr)}\cz_{T}'+ \frac{c_R^2}{c_T^2}\Bigl(\omega^2-q^2 \frac{c_T^2}{c_X^2}\Bigr)\cz_{T}=0\,.
\end{equation}
With the ansatz (\ref{ansatztensorial}) one can check that the only non-zero term in the solution normalized to one at the horizon up to first order in $\lambda_{hyd}$ is $\cz_{T}^{0,0}=1$.
At second order in $\lambda_{hyd}$ the solution is too lengthy to be reported here but straightforward to obtain.

\subsection*{Scalar perturbations}

The equations for the scalar gauge invariant fluctuations are the relevant ones for the sound channel, giving the dispersion relation in (\ref{vecdiff2}).
They are just combinations of the equations in \cite{benincasa}:\footnote{$H_{aa}$ in \cite{benincasa} corresponds to our $H_{\pperp}$.}
\begin{eqnarray}
0 &=& 2 H_{zz}^{EOM}+4\frac{q}{\omega}H_{tz}^{EOM}-\Bigl(1-\frac{q^2}{\omega^2}\frac{c_T'c_T}{c_X' c_X} \Bigr)H_{aa}^{EOM}+2\frac{q^2}{\omega^2}\frac{c_T^2}{c_X^2}  H_{tt}^{EOM}+\nonumber\\
&& + \Bigl(\frac{\omega}{q}\frac{c_X^2}{c_T^2}\Xi+\frac{8}{\omega}\log'{\frac{c_X}{c_T}} \Bigr)H_{rt}^{EOM}+\Xi H_{rz}^{EOM}\, ,\\
0 &=& \phi^{EOM}-\frac{\phi_B'}{\log'{c_X^4}}H_{aa}^{EOM}+\frac{\omega c_X^2[c_X c_X'\phi_B''+\phi_B'(c_X'^2-c_X c_X'')]}{c_X'(q^2 c_T c_T' c_X+2q^2 c_T^2c_X'-3\omega^2c_X^2c_X')}\Bigl( H_{rt}^{EOM}+\frac{q c_T^2}{\omega c_X^2}H_{rz}^{EOM}\Bigr)\, .\nonumber
\end{eqnarray}
In these expressions, $\phi$ represents each of the scalars $f, w, \Phi$ (the form of their equation is the same) and $\phi_B$ their background value. 
Moreover, the notation $\phi^{EOM}$ (and $H_{zz}^{EOM}$ and so on) stands for the corresponding equation for the scalar (and the fluctuation $H_{zz}$ and so on) in section 3 of \cite{benincasa}. For the coefficient $\Xi$ we have
\begin{eqnarray}
\Xi &=&-\frac{16q r_h^4(2q^2-3\omega^2)}{r\omega^2[q^2(r_h^4-3r^4)+3r^4\omega^2]}-\frac{4q^3 r_h^4(q^2-\omega^2)(r^4-r_h^4)}{r\omega^2[q^2(r_h^4-3r^4)+3r^4\omega^2]^2}\epsilon_*^2\, .
\end{eqnarray}

We give here the solution to the non-zero fluctuations entering in the sound channel, satisfying the normalization at the horizon and Dirichlet conditions at the boundary in the case of $\cz_S$ 
 \begin{eqnarray*}
 \cz_S^{0,0} & = & \frac{1}{\rho^{4}}\,,\qquad \cz_\cph^{0,2}=\frac{\log{\rho}}{12(1-\rho^4)} \,,\rc
 \cz_\cph^{1,2} &=& \frac{i}{144\sqrt{3}(1-\rho^4)} \Bigl[ \pi^2(\rho^4-1) +24 (\rho^4-1) \log^2{\rho} -12 \log{\rho}\Bigl(4+ (\rho^4-1)  \log{(1+i\rho)}+\nonumber \\
 && (\rho^4-1)  \log{[i(i+\rho)(\rho^2-1)]}  \Bigr) -3(\rho^4-1)Li_2(\rho^4)  \Bigl]\,,
\end{eqnarray*}
where $\rho\equiv r/r_h$. We do not report the expressions for all the $\qn^2$ coefficients of the solutions because of their very lengthy form.

\subsection*{Vectorial perturbations
}

The equation in this channel reads
\begin{equation}
\cz_{V}''+\Bigl[ \log'{\Bigl(\frac{c_X^5}{c_T  c_R}\Bigr)}-\log'{\Bigl(\frac{c_X^2}{c_T^2}\Bigr)}\Bigl(1-\frac{q^2}{\omega^2} \frac{c_T^2}{c_X^2}\Bigr)^{-1}\Bigr] \cz_{V}'+ \frac{c_R^2}{c_T^2}\Bigl(\omega^2-q^2 \frac{c_T^2}{c_X^2}\Bigr)\cz_{V}=0\,.
\end{equation}
For the vectorial fluctuations we can solve order by order with the scaling\footnote{Scaling the frequency also with $\wn\to \lambda_{hyd}\wn$ gives the same answer, as is the case in \cite{Kovtun:2005ev}.} $\wn\to \lambda_{hyd}^2 \wn$,  $\qn\to \lambda_{hyd} \qn$, imposing regularity at the horizon. The result is
\be
\cz_V = C_V \left(1-  \frac{r_h^4}{r^4}\right)^{-i\frac{\wn T_{0}}{2 T}}  \left[ \frac{r_h^4}{r^4}+\left( 1- i \frac{\qn^2}{\wn} \right)  \left(1-\frac{r_h^4}{r^4} \right) (1+\epsilon_*+\epsilon_*^2) \right] + {\cal O}(\wn,\qn^2) \, .
\ee
From Dirichlet conditions at the boundary $r_*\to\infty$ we can read off the shear viscosity from the dispersion relation
\be
\omega = -i\,\frac{\eta}{sT} q^2 + {\cal O}(q^3)\, .
\ee
This calculation is summarized in the membrane paradigm formula given in \cite{Kovtun:2003wp}, and it gives the well-stated ratio $\eta/s = 1/(4\pi)$ with corrections in powers of $r_h/r_* \to0$.


\end{document}